\documentclass[prb,amsmath,amssymb,superscriptaddress,twocolumn]{revtex4}
\usepackage{float}
\usepackage{amsmath}
\usepackage{amssymb}
\usepackage{amsfonts}
\usepackage{euscript}
\usepackage{enumerate}
\usepackage{hhline}
\usepackage{pslatex}
\usepackage{tabularx}
\usepackage[usenames,dvipsnames]{xcolor}

\usepackage{graphicx}

\usepackage{dcolumn}
\usepackage{bm}
\usepackage[sort&compress]{natbib}

\usepackage{lipsum}
\usepackage{textcomp, gensymb}

\makeatletter
\renewcommand{\@biblabel}[1]{#1. }
\renewcommand{\@dotsep}{500}
\renewcommand{\@pnumwidth}{0em}
\renewcommand{\l@figure}[2]{
\@dottedtocline{1}{1.5em}{2em}{Figure #1}{}\vspace{15pt}}

\usepackage[normalem]{ulem}
\usepackage{upgreek}

\begin{document}

\title{Parasitic loss in microring-waveguide coupling and its impact on wideband nonlinear photonics}

\author{Yi Sun}
\affiliation{Microsystems and Nanotechnology Division, Physical Measurement Laboratory, National Institute of Standards and Technology, Gaithersburg, MD 20899, USA}
\affiliation{Joint Quantum Institute, NIST/University of Maryland, College Park, MD 20742, USA}

\author{Daniel Pimbi}
\affiliation{Microsystems and Nanotechnology Division, Physical Measurement Laboratory, National Institute of Standards and Technology, Gaithersburg, MD 20899, USA}
\affiliation{Joint Quantum Institute, NIST/University of Maryland, College Park, MD 20742, USA}

\author{Xiyuan Lu}\email{xnl9@umd.edu}
\affiliation{Microsystems and Nanotechnology Division, Physical Measurement Laboratory, National Institute of Standards and Technology, Gaithersburg, MD 20899, USA}
\affiliation{Joint Quantum Institute, NIST/University of Maryland, College Park, MD 20742, USA}

\author{Jordan Stone}
\affiliation{Microsystems and Nanotechnology Division, Physical Measurement Laboratory, National Institute of Standards and Technology, Gaithersburg, MD 20899, USA}
\affiliation{Joint Quantum Institute, NIST/University of Maryland, College Park, MD 20742, USA}

\author{Junyeob Song}
\affiliation{Microsystems and Nanotechnology Division, Physical Measurement Laboratory, National Institute of Standards and Technology, Gaithersburg, MD 20899, USA}

\author{Zhimin Shi}
\affiliation{Reality Labs Research, Meta, Redmond, Washington 98052, USA}

\author{Kartik Srinivasan}\email{kartik.srinivasan@nist.gov}
\affiliation{Microsystems and Nanotechnology Division, Physical Measurement Laboratory, National Institute of Standards and Technology, Gaithersburg, MD 20899, USA}
\affiliation{Joint Quantum Institute, NIST/University of Maryland, College Park, MD 20742, USA}

\date{\today}

\begin{abstract}
     \noindent Microring resonators enable the enhancement of nonlinear frequency mixing processes, generating output fields at frequencies that widely differ from the inputs, in some cases by more than an octave. The efficiency of such devices depends on effective in- and out-coupling between access waveguides and the microrings at these widely separated frequencies. One successful approach is to separate the coupling task across multiple waveguides, with a cutoff waveguide (a waveguide that does not support guided modes above a certain wavelength) being judiciously used to prevent unwanted excessive overcoupling at low frequencies. Here, we examine how such a cutoff waveguide can still induce parasitic loss in the coupling region of a microring resonator, thereby impacting nonlinear device performance. We verified this parasitic loss channel through both experiment and simulation, showing that a waveguide optimized for 532 nm (visible) and 780 nm (near-infrared), while nominally cut off at 1550 nm, can still introduce significant parasitic loss at telecom wavelengths. This is studied in the context of visible-telecom optical parametric oscillation, where the excess parasitic loss can be strong enough to prevent threshold from being reached. Our finding elucidates a major challenge for wideband integrated nonlinear photonics processes when efficient coupling of widely-separated frequencies is needed.
\end{abstract}

\maketitle
\section{Introduction}
\noindent
Nonlinear nanophotonics enables chip-integrated coherent light generation at frequencies outside of those available through conventional optical gain media, such as direct band gap gain in semiconductors. For high efficiency, such processes commonly rely on microresonators to enhance the effective strength of the nonlinear interaction~\cite{Kippenberg_Science_2018,Poon_AOP_2024,Lu_NatPhoton_2024_review}. Nonlinear conversion into the visible or within the visible is possible through a variety of second- and third-order nonlinear processes, such as Kerr microcomb generation~\cite{Obrzud_OL_2019_visible_comb,Zhao_Optica_2020,Moille_LPR_2025_visible_comb,Corato_OE_2025_visible_comb,Li_LPR_2025_visible_comb_early}, second- and third-harmonic generation (SHG and THG)~\cite{Carmon_NatPhys_2007_Vahala_THG_green,Lin_PRL_2016,Surya_Optica_2018_THG_HT,Surya_OptLett_2018,Chang_APLPhoton_2019_SHG_GaAs,Lu_NaturePhotonics_2020_sin_shg,Hu_SciAdv_2022_Bres_SHG,Nitiss_OE_2023_sin_SHG_Bres,Li_Optica_2023_oxide_multipleWG}, optical parametric oscillation (OPO)~\cite{Lu_Optica_2019C,Lu_Optica_2020_OPO_visible,Domeneguetti_Optica_2021,Jordan_APLPhotonics_2022_OPO_efficiency,Ng_OE_2023_MgF2_two-mode_OPO,Sun_LSA_2024_gOPO}, stimulated four-wave mixing (StFWM)~\cite{Lu_NatPhoton_2019B,Ramelwow_StFWM}, stimulated Brillouin scattering~\cite{Chauhan_NatCommun_2021_Blumenthal_SBS}, and FWM-Bragg scattering (FWM-BS)~\cite{Li_NatPhoton_2016}. The successful implementation of each of these nonlinear processes is based on management of microresonator nonlinearity, quality factor, dispersion, and waveguide coupling.

\begin{figure}[t!]
\centering\includegraphics[width=0.95\linewidth]{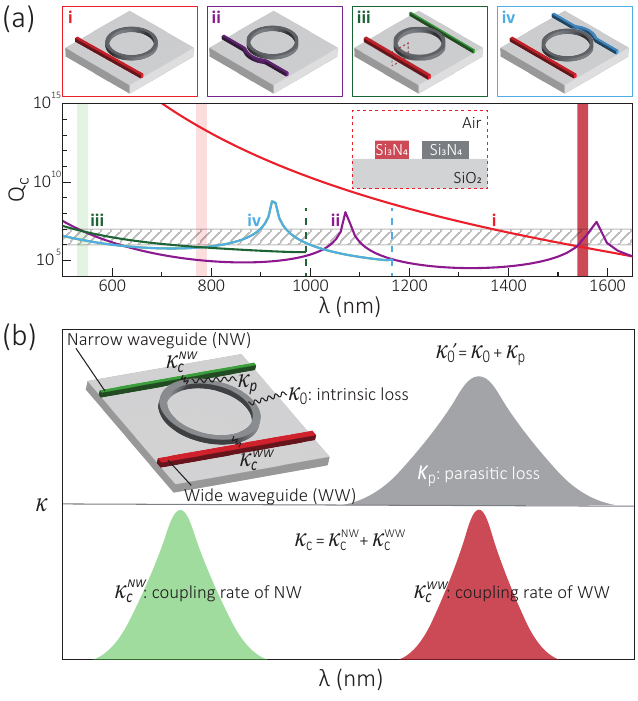}
\caption{\small \textbf{(a)} Four possible coupling schemes (top) for wideband, air-clad nonlinear integrated photonics that are designed to optimize coupling quality factor ($Q_c$) across visible (532 nm), near-infrared (780 nm), and telecom (1550 nm) wavelength bands highlighted in light green, pink and red, respectively (bottom). The inset shows a schematic of the coupling region between the waveguide (red) and the microring resonator (dark gray). The hatched gray rectangular area in the representative $Q_c$ spectra denotes the design target for critical coupling with $Q_c=10^6$~to~$10^7$. \textbf{i.} Single straight waveguide (red): critically coupled at 1550~nm, but severely undercoupled at shorter wavelengths; \textbf{ii.} Single pulley waveguide (purple): broadband coupling with a fabrication-sensitive pulley anti-resonance preventing overcoupling at 1550~nm. Despite this design effort, $Q_c$ varies by two orders between the three targeted bands; \textbf{iii.} Two separate straight waveguides: one narrow, small-gap waveguide (green) is optimized for visible and near-infrared with a cutoff wavelength near 1000 nm, ensuring it only guides shorter wavelengths. The other wide, large-gap waveguide (red) is optimized for critical coupling at 1550~nm; \textbf{iv.} One straight waveguide combined with one pulley waveguide: similar to scheme \textbf{iii}, the straight waveguide (red) is designed for critical coupling at 1550~nm but now the narrow pulley waveguide  (blue), with a cutoff wavelength near 1200 nm, has more optimal coupling at 532 nm and 780 nm. These two waveguides thus, in principle, cover all three bands optimally. \textbf{(b)} Detailed explanation of loss channels for scheme \textbf{iii} in \textbf{(a)}: the intrinsic loss ($\kappa_0$), dominated by material absorption and microring sidewall roughness scattering, exhibits weak wavelength dependence (varying within a factor of 3). For efficient coupling, the coupling rate ($\kappa_c^{NW}$ and $\kappa_c^{WW}$ for the narrow and wide waveguides, respectively) is typically designed to match $\kappa_0$, with slight under- or over-coupling depending on application needs. Here, we identify a parasitic loss channel ($\kappa_\text{p}$), induced by the narrow waveguide, which elevates the effective intrinsic loss ($\kappa_0'$ = $\kappa_0$ + $\kappa_\text{p}$) for longer wavelengths. Such loss is often overlooked but can have a strong impact on performance.}
\label{Fig1}
\end{figure}

Microresonator-waveguide coupling is especially challenging when optimizing for multiple, widely-separated frequencies. Evanescent coupling depends on modal overlap, phase mismatch, and interaction length, and in general, the coupling rate increases as frequency decreases (wavelength increases), due to increased modal overlap. For microrings, point coupling, where the waveguide is tangent to the ring, can exhibit a coupling rate variation of a few orders-of-magnitude across an octave~\cite{Moille_OptLett_2019_comb_pulley}, as depicted in coupling scheme \textbf{i} in Fig.~\ref{Fig1}(a), with the geometry shown on top and a typical coupling quality factor ($Q_\text{c}$) spectrum shown on bottom. Thus, a waveguide designed to be critically coupled at short wavelengths results in strong overcoupling at long wavelengths. Pulley coupling [scheme \textbf{ii} in Fig.~\ref{Fig1}(a)], where the waveguide is wrapped around a portion of the ring, can reduce such coupling dispersion, though even optimized designs can exhibit more than one order of magnitude variation when bandwidths extend beyond an octave~\cite{Moille_OptLett_2019_comb_pulley, Moille_NatComm_2021_fwmbs_comb}. To mitigate strong overcoupling in targeted spectral bands, the pulley coupling anti-resonances, in which local regions of the spectrum exhibit pronounced reduced coupling, can be carefully engineered, but is often highly sensitive to dimensional variations. On the other hand, unoptimized coupling strongly impacts the performance of wideband nonlinear nanophotonics devices, for example, in term of conversion efficiency, output power, and other metrics.

A successful solution to this wideband coupling challenge is to take advantage of cutoff in waveguides with asymmetric cladding, which can be realized through a silicon dioxide (SiO$_2$) layer underneath the waveguide and air surrounding on all other sides, as shown in Fig.~\ref{Fig1}(a). For a given waveguide geometry, there exists a cutoff wavelength, above which guided modes are not supported~\cite{Yariv_Photonics_2007}. This property doesn't exist for symmetrically clad rectangular waveguides, resulting in $Q$ degradation at long wavelengths caused by parasitic loss, even when careful design is employed~\cite{Li_Optica_2023_oxide_multipleWG}. Using cutoff, the coupling task for wideband nonlinear photonics can be split between two (or more) waveguides, with each responsible for different spectral regions. Typically, a narrow width, small gap waveguide addresses shorter wavelengths (and is cutoff at long wavelengths), while a wider width, large gap waveguide addresses longer wavelengths (with negligible coupling at short wavelengths due to the large gap). Representative cases where the narrow width waveguide is point coupled [\textbf{iii} in Fig.~\ref{Fig1}(a)] or pulley coupled [\textbf{iv} in Fig.~\ref{Fig1}(a)] to the microring, and the wider width waveguide is point coupled to the microring in the same fashion as in \textbf{i} in Fig.~\ref{Fig1}(a), have been used for octave-spanning microcombs~\cite{Li_CLEO_2015_OctaveComb,Briles_OptLett_2018_interlocking}, highly non-degenerate entangled photon pair sources~\cite{Lu_NatPhys_2019A}, and telecom-to-visible spectral translation~\cite{Lu_NatPhoton_2019B}. In addition to the coupling benefits, dispersion wise, the use of asymmetrically-clad structures naturally dovetails with visible wavelength dispersion engineering~\cite{Lu_Optica_2020_OPO_visible,Moille_LPR_2025_visible_comb}, where a top air cladding eliminates the normal dispersion of the typical SiO$_2$ top cladding, which would otherwise require compensation in the resonator design.

A crucial assumption underlying the above strategy is that the cutoff waveguide has no impact on the resonator for sufficiently long wavelengths. However, in this paper, we report that this plausible assumption needs to be examined and executed with care: it turns out that even a waveguide satisfying the cutoff condition can still induce substantial parasitic loss ($\kappa_\text{p}$) in the microring, as depicted in Fig.~\ref{Fig1}(b), either due to scattering or coupling to leaky modes. We quantitatively study the impact of $\kappa_\text{p}$ in the context of wideband Kerr OPO. Such OPOs enable the generation of visible light from a near-infrared pump~\cite{Lu_Optica_2020_OPO_visible,Domeneguetti_Optica_2021}, with wavelengths across the full green gap spectral region recently been reported~\cite{Sun_LSA_2024_gOPO}. Our study highlights how the previously neglected $\kappa_\text{p}$ contribution from a cutoff waveguide can limit extension of OPO spectral coverage deeper into the visible. In particular, we show that degradation of the $Q$ of the long-wavelength idler mode inhibits OPO threshold from being reached, thereby setting a limit on the longest (and therefore shortest) wavelengths achievable in a given OPO design. Our results are vital in understanding how to design nonlinear nanophotonic devices to realize efficient conversion across the full range of wavelengths allowed by the resonator dispersion and material transparency.

\begin{figure}[t!]
\centering\includegraphics[width=0.95\linewidth]
{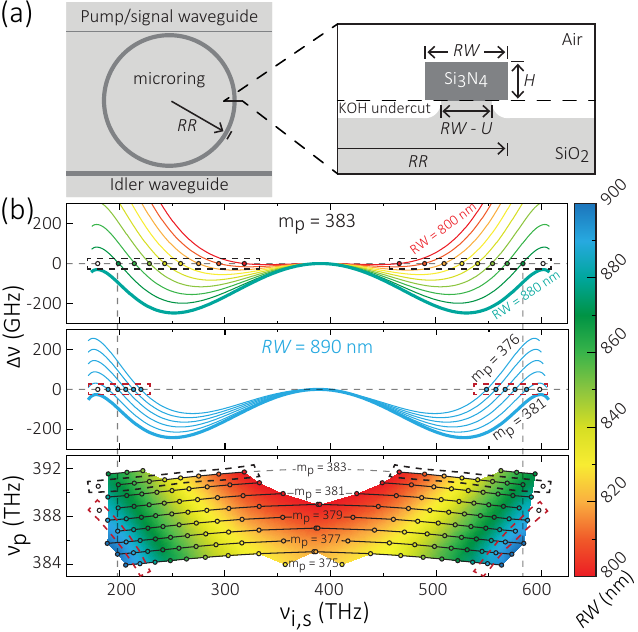}
\caption{\small \textbf{(a)} Schematic of a Kerr optical parametric oscillator (OPO) device with a zoomed-in cross-sectional view. The microring resonator features two waveguides for coupling the pump/signal and idler, respectively. The undercut structure is symmetrically etched using heated potassium hydroxide treatment. Key geometric parameters include the outer ring radius ($RR$), ring width ($RW$),  Si$_3$N$_4$ thickness ($H$), and total undercut ($U$). \textbf{(b)} Simulated spectral coverage for OPO devices ($RR$ = 25~{\textmu}m, $H = 600$~nm, $U = 140$~nm, $RW = 800$~nm~to~$900$~nm). Top panel: Frequency mismatch spectra ($\Delta\nu = \nu_{\rm{s}} + \nu_{\rm{i}} - 2\nu_{\rm{p}}$) for $RW = 800$~nm~to~880~nm with fixed $m_{\rm{p}} = 383$. Middle panel: Spectra for $RW = 890$~nm with $m_{\rm{p}} = 376$~to~381. Solid circles mark zero crossings, which predicted $\nu_{\rm{s}}$ and $\nu_{\rm{i}}$ frequencies with empty circles indicating configurations in which OPO is not supported. Bottom panel: Spectral map showing OPO coverage from $\approx190$~THz to $\approx590$~THz. Color scale denotes $RW$ and odd $m_{\rm{p}}$ values are labeled. Dashed black and red boxes link to cases in the top and middle panels, respectively. Vertical dashed lines highlight degenerate $\nu_{\rm{s}}$/$\nu_{\rm{i}}$ generation via multiple $RW$ and $m_{\rm{p}}$ combinations.}
\label{Fig2}
\end{figure}

\section{Results and discussions}
Out of various wideband nonlinear optical processes, we choose Kerr OPO as it is a highly informative tool to unveil the impact of the cutoff coupling waveguide on nonlinear light generation. The versatility of OPO lies in its ability to generate light across a wide range of signal and idler frequencies ($\nu_{\rm{s}}$ and $\nu_{\rm{i}}$) by varying the pump frequency ($\nu_{\rm{p}}$), with a change of $\approx5$~THz enough to shift the output frequencies by $>120$~THz~\cite{Lu_Optica_2020_OPO_visible,Sun_LSA_2024_gOPO}. Moreover, dispersion engineering by adjusting the ring width ($RW$) between devices on the same chip provides further tuning of the output frequencies~\cite{Lu_Optica_2019C,Lu_Optica_2020_OPO_visible,Domeneguetti_Optica_2021,Jordan_APLPhotonics_2022_OPO_efficiency,Sun_LSA_2024_gOPO}. These characteristics enable a detailed analysis of the onset of performance degradation in terms of the widest OPO separation achievable at a given pump power, as the OPO threshold is inversely proportional to the $Q$s of all three modes involved~\cite{Lu_Optica_2019C}. This means that a sufficiently large reduction in $Q$ as the targeted signal/idler bands become increasingly spectrally separated from the pump -- for example, due to excess parasitic loss induced by the cutoff waveguide at long wavelengths -- would prevent threshold from being reached. Such analysis would be challenging with other nonlinear processes like SHG and THG, which have no threshold power and generally lack the capacity to probe a broad set of output wavelengths within a single device. In addition, OPO, as an oscillation process seeded by vacuum fluctuations, is constrained by its threshold power, while the complementary process of StFWM, where a seed laser is injected along with the pump, is thresholdless. They can thus be used together to validate whether the inability to reach threshold is due to resonator dispersion or loss~\cite{Lu_NatPhoton_2019B}.

We utilize high-$Q$ silicon nitride (Si$_3$N$_4$) microring resonators as a platform for our widely-separated OPO devices. The chip uses an underlying SiO$_2$ cladding and top side air cladding with all key parameters depicted in Fig.~\ref{Fig2}(a). Figure.~\ref{Fig2}(b) shows the simulated spectral coverage for such OPO devices with $RR = 25$~{\textmu}m, $H = 600$~nm, $U = 140$~nm, and $RW$ ranging from $800$~nm to $900$~nm. The undercut $U$, achieved by heated potassium hydroxide etching, helps to enhance the robustness of OPO generation~\cite{Sun_LSA_2024_gOPO}. The pump wavelength is constrained from $765$~nm to $781$~nm, matching the tuning range of the external cavity diode laser (ECDL) pump laser, corresponding to the pump mode azimuthal number ($m_{\rm{p}}$) ranging from $375$ to $384$ for the fundamental transverse electric mode ($\text{TE}_{0}$) with the microring resonator geometries mention above. Simulations in Fig.~\ref{Fig2}(b) are performed through eigenmode calculations of the microrings using the finite-element method. For Kerr OPOs, a pair of $\nu_{\rm{i}}$ and $\nu_{\rm{s}}$ that are widely separated from $\nu_{\rm{p}}$ can be achieved when the pump dispersion is normal but close to zero~\cite{Lu_Optica_2019C}, and the frequency mismatch, defined as $\Delta\nu = \nu_{\rm{s}}+\nu_{\rm{i}}-2\nu_{\rm{p}}$, has a small but positive value ($\Delta\nu~\gtrsim~0$). Two sets of frequency mismatch spectra are shown in the top and middle panels in Fig.~\ref{Fig2}(b). The top panel shows frequency mismatch spectra for nine devices with $RW$ varying from $800$~nm to $880$~nm at a fixed $m_{\rm{p}}~=~383$, while the middle panel presents the spectra for a single device with $RW$ = $890$~nm while $m_{\rm{p}}$ ranges from $376$ to $381$. The solid circles are the zero-crossing points, which representing the prospective $\nu_{\rm{i}}$ and $\nu_{\rm{s}}$ for different $RW$ and $m_{\rm{p}}$ combinations. The two bold curves don't have zero-crossing points, indicating that OPOs are not supported in those cases. $m_{\rm{p}}$ and $RW$ thus serve as two tuning knobs for OPOs: When $m_{\rm{p}}$ is fixed at 383, as $RW$ is increased from $800$~nm to $870$~nm, $\nu_{\rm{i}}$ ($\nu_{\rm{s}}$) could cover the spectral range between $\approx318$~THz ($\approx465$~THz) to $\approx199$~THz ($\approx582$~THz); when $RW$ is fixed at $890$~nm, as $m_{\rm{p}}$ is increased from $376$ to $381$, $\nu_{\rm{i}}$ ($\nu_{\rm{s}}$) could cover the spectral range between $\approx219$~THz ($\approx549$~THz) to $\approx189$~THz ($\approx586$~THz). The bottom panel in Fig.~\ref{Fig2}(b) shows the simulated spectral coverage of OPOs after exhausting all possible combinations of $m_{\rm{p}}$ and $RW$. In theory, these OPOs could generate any light between $\approx190$~THz to $\approx590$~THz.

In addition, the two vertical dashed lines connecting three panels in Fig.~\ref{Fig2}(b) demonstrate that several combinations of $RW$ and $m_{\rm{p}}$ [parameterized as ($RW$, $m_{\rm{p}}$)] can be employed to generate a specific $\nu_{\rm{i}}$ or $\nu_{\rm{s}}$. For example, combinations of ($870$~nm, $383$), ($880$~nm, $381$), ($890$~nm, $379$), and ($900$~nm, $377$) could generate $\nu_{\rm{i}}$~$\approx200$~THz, and other combinations of ($870$~nm, $383$) and ($900$, $378$) could generate $\nu_{\rm{s}}$~$\approx582$~THz. This ability to reach the same $\nu_{\rm{i}}$ or $\nu_{\rm{s}}$ through multiple $RW$ and $m_{\rm{p}}$ configurations provides a critical diagnostic capability: If a broader $\nu_{\rm{i}}$-$\nu_{\rm{s}}$ span, as predicted by dispersion simulations, remains unattainable despite testing all viable $RW$ and $m_{\rm{p}}$ combinations, the limitation must stem from a parameter outside those governing dispersion — specifically, as we will discuss later, the cutoff waveguide width.
 
\begin{figure*}[t!]
\centering\includegraphics[width=0.95\linewidth]
{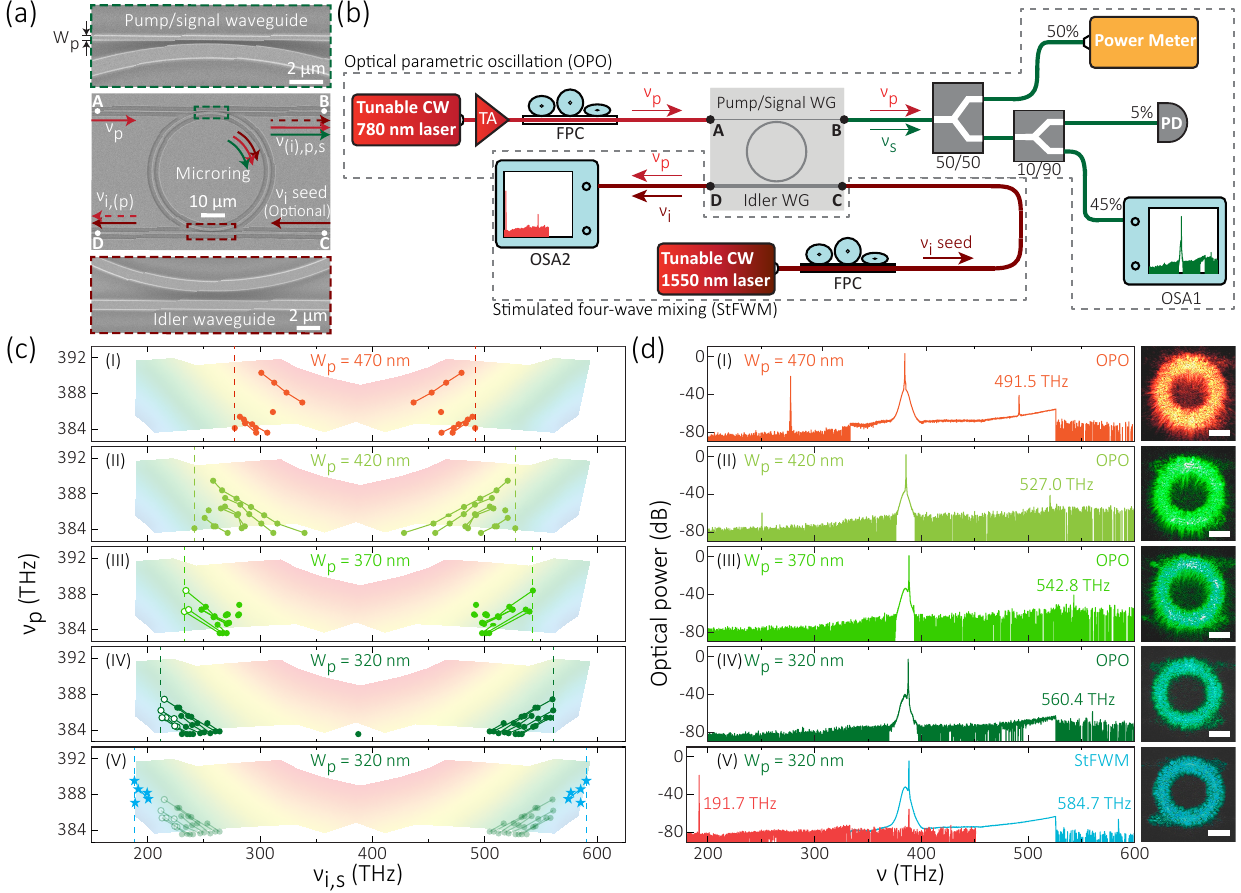}
\caption{\small \textbf{(a)} Scanning electron microscope (SEM) images of a Kerr OPO microring resonator, with zoomed-in views of pump/signal and idler coupling regions. The width of the pump/signal waveguide is defined as $W_\text{p}$. $\nu_{\rm{p}}$ is injected through port A, with the residual pump and $\nu_{\rm{s}}$ collected at port B. In cases where $\nu_{\rm{i}}$ is cutoff through the pump/signal waveguide, it can be detected at port D together with residual pump. For stimulated four-wave mixing (StFWM), a seed at $\nu_{\rm{i}}$ is injected at port C at the same time as the pump in port A. \textbf{b} Experimental setup for OPO and StFWM. TA: Tapered amplifier, FPC: Fiber polarization controller, OSA: Optical spectrum analyzer, PD: Photodiode. \textbf{(c)} Experimentally observed $\nu_{\rm{s}}$ and $\nu_{\rm{i}}$ for OPO (circles, I–V) and StFWM (stars, V) with different $W_\text{p}$. Solid circles: $\nu_{\rm{s}}$ and $\nu_{\rm{i}}$ directly obtained from OSA1; empty circles: $\nu_{\rm{i}}$ calculated via energy conservation from measured $\nu_{\rm{p}}$ and $\nu_{\rm{s}}$. Stars: $\nu_{\rm{s}}$ measured at OSA1 and $\nu_{\rm{i}}$ measured at OSA2. Simulated spectral coverage from Fig.~\ref{Fig2}(b) (shadowed in the background for comparison) closely matches experimental data, using identical geometries. Vertical dashed lines mark maximum \(\nu_{\rm{i}} - \nu_{\rm{s}}\) spans for each $W_\text{p}$. \textbf{(d)} Widest OPO (I – IV) and StFWM (V) spectra in \textbf{(c)}. Here, $0$~dB is referenced to $1$~mW, i.e., dBm. Idler cutoff through the pump/signal waveguide occurs in III, IV, and V collecting no idler in OSA1, while spectrum in V combines the spectra recorded from OSA1 and OSA2. The right panel shows the optical microscope images of scattered signal light with 20~{\textmu}m scale bars.}  
\label{Fig3}
\end{figure*}

In order to experimentally investigate the influence of $\kappa_\text{p}$ on the idler due to the narrow coupling waveguide, we fabricate these OPO devices with dimensions as mentioned above while also varying the width of the pump/signal waveguide ($W_\text{p}$), thereby varying the wavelength at which long-wavelength cutoff occurs. The device design is created with the NIST Center for Nanoscale Science and Technology’s Nanolithography Toolbox~\cite{Balram_JResNatlInst_2018}. A silicon wafer is thermally oxidized to produce a $3$~{\textmu}m thick SiO$_2$ layer, followed by deposition of a 600~nm thick Si$_3$N$_4$ film using low-pressure chemical vapor deposition. Layer thicknesses and wavelength-dependent refractive indices are determined via spectroscopic ellipsometry and modeled with an extended Sellmeier equation. The wafer is then diced into small pieces for patterning. Electron beam lithography with positive tone resist defines the patterns, which are transferred to the Si$_3$N$_4$ via oxygen and fluoride-based reactive-ion etching. Post-etch chemical cleaning removes residual polymer, and the device is annealed at $1100$~$\degree$C for 4 hours under nitrogen. An oxide lift-off process selectively air-clads the devices while preserving SiO$_2$ cladding on input/output waveguides for efficient edge coupling. Microring undercuts are formed via $70$~$\degree$C KOH etching, with lateral etching rates calculated from vertical rate measurements and validated by cross-sectional scanning electron microscope (SEM) imaging. The chip is diced, polished, and prepared for lensed-fiber coupling. 

Figure~\ref{Fig3}(a) shows the SEM images of the fabricated devices. Two straight waveguides are used to inject and extract light in and out of the OPOs, following scheme \textbf{iii} from Fig.~\ref{Fig1}(a). The narrow waveguide, of width $W_\text{p}$ and with a small coupling gap, is designed to in-couple the pump through port A and out-couple the pump and signal light through port B, and is cut-off in the idler band. The second (wide) waveguide with a large coupling gap is designed to out-couple the idler light at long wavelengths through port D. The zoomed-in SEM images show the coupling regions for these two waveguides. The measurement setup is shown in Figure~\ref{Fig3}(b). An amplified ECDL with tunable range between $765$~nm to $781$~nm is injected from port A through lensed fiber after fiber polarization controller. The on-chip pump power is limited $\approx$~100~mW with a calibrated insertion loss $\approx$~3 dB per facet for the pump. The device $RW$ and pump mode selected act as two tuning parameters, as illustrated previously in Figure~\ref{Fig2}(b). OPO can be observed in devices with appropriate $RW$ configurations when the on-chip pump power exceeds the threshold while the pump wavelength is scanned from shorter to longer wavelengths, provided dispersion conditions are properly engineered. When OPO is generated, $\nu_{\rm{s}}$, $\nu_{\rm{p}}$, and $\nu_{\rm{i}}$ (when shorter than the cutoff wavelength for the pump/signal waveguide) can be recorded from OSA1 through port B. OSA2 through port D can be used to record $\nu_{\rm{i}}$ and part of $\nu_{\rm{p}}$ as the wider idler waveguide is designed for critically coupling $\nu_{\rm{i}}$ but undercoupling for $\nu_{\rm{p}}$ and $\nu_{\rm{s}}$. Another tunable laser at telecom wavelengths near $\nu_{\rm{i}}$ can be injected from port C (the wide waveguide input) in addition to $\nu_{\rm{p}}$ injection at port A for StFWM experiments, which can be used to validate the dispersion.

Figure~\ref{Fig3}(c)[(I) - (IV)] show the experimentally observed OPO $\nu_{\rm{i}}$ and $\nu_{\rm{s}}$ distributions for different $W_\text{p}$ values. Data points within the same connected line are from a single device with the same $RW$. Simulated spectral coverage in the bottom panel of Fig.~\ref{Fig2}(b) is shadowed in the background, and shows a good consistency between the simulation and experiment. Figure~\ref{Fig3}(d)[(I) - (IV)] 
display the widest achievable OPO spectrum for different $W_\text{p}$, with the right panel showing representative optical microscope images of scattered light of the generated visible signal. This plot clearly shows that, for a given $W_\text{p}$, there is a $\nu_{\rm{i}}$-$\nu_{\rm{s}}$ span limit (marked with dashed lines), with multiple OPO devices with different $RW$ reaching this maximum OPO span at different $\nu_{\rm{p}}$. The widest $\nu_{\rm{i}}$-$\nu_{\rm{s}}$ span increases from $\approx210$~THz to $\approx350$~THz as $W_\text{p}$ decreases from $470$~nm to $320$~nm, as demonstrated from (I) to (IV). Clearly, $W_\text{p}$ significantly impacts the maximum OPO span, even if the resonator dispersion and pump mode suggest that wider OPO is achievable.

To that end, we perform StFWM experiments to confirm that our OPO span is not limited by an inability to achieve phase- and frequency-matching. For the same devices with $W_\text{p}$ = 320~nm, when injecting a seed laser near $\nu_{\rm{i}}$ from port C in addition to the near-infrared pump from port A, we are able to see extended $\nu_{\rm{i}}$-$\nu_{\rm{s}}$ spans up to $\approx390$~THz, thereby achieving the edge of the simulated spectral coverage, as shown in Fig.~\ref{Fig3}(c)(V). Figure~\ref{Fig3}(d)(V) displays a typical StFWM spectrum created by combining the spectra recorded from OSA1 and OSA2, while the right panel shows the generated representative cyan color scattered from the microring. Compared to OPO, StFWM is a significantly more efficient process with no threshold power requirement~\cite{Lu_NatPhoton_2019B}. This extension of the  $\nu_{\rm{i}}$-$\nu_{\rm{s}}$ span for the same set of devices under StFWM demonstrates that the resonator dispersion is not likely to be the limitation on the OPO span in (IV), and suggests that $Q$ degradation may instead be the cause.

\begin{figure}[t!]
\centering\includegraphics[width=0.95\linewidth]{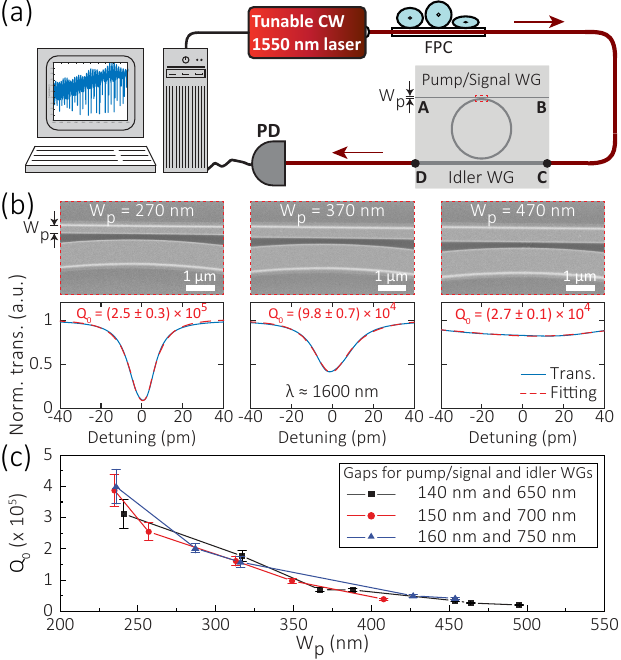}
\caption{\small \textbf{(a)} Setup for telecom wavelength transmission measurements through the idler waveguide and in the presence of a varying width pump/signal waveguide. \textbf{(b)} Top-view SEM images for the pump/signal waveguide coupling region with red dashed box in \textbf{a} for $W_\text{p}$ = $270$~nm, $370$~nm, and $470$~nm. The bottom panels show corresponding Q measurements and fits near $1600$~nm for these $W_\text{p}$ values. \textbf{(c)} Summary of the intrinsic quality factor ($Q_0$) for modes near 1600~nm as a function of $W_\text{p}$ with three different coupling gap configurations, which clearly shows that $Q_0$ at the telecom wavelength is significantly degraded as $W_\text{p}$ increases. The uncertainties in \textbf{(b)} and error bars in \textbf{(c)} are one standard deviation values from the fits.}
\label{Fig4}
\end{figure}

To investigate the reason for OPO performance degradation with increasing $W_\text{p}$, we measure the intrinsic quality factor ($Q_0$) near $1600$~nm using the idler waveguide as shown in Fig.~\ref{Fig4}(a). These devices feature the same geometric parameters including the $RR$ of 25~{\textmu}m, $H$ of $600$~nm, and $RW$ of $850$~nm, with the only variation being in $W_\text{p}$, which ranges from $235$~nm to $495$~nm. This narrow pump/signal waveguide is designed to be cutoff at telecom wavelengths for all the $W_\text{p}$ values. Figure~\ref{Fig4}(b) shows SEM images for the coupling regions, transmission measurements, and the fits to extract $Q_0$ at $1600$~nm when $W_\text{p}$ = $270$~nm, $370$~nm, and $470$~nm. We find that $Q_0$ decreases from $\approx2.5\times10^5$ to $\approx2.7\times10^4$ as $W_\text{p}$ increases from $270$~nm to $470$~nm. Figure~\ref{Fig4}(c) summarizes the $Q_0$ trends with different $W_\text{p}$ values under three different coupling gap configurations, and clearly show a substantial decrease of $Q_0$ with increasing $W_\text{p}$ in all cases. This significant reduction in $Q_0$ for a prospective idler mode elevates its OPO threshold~\cite{Lu_Optica_2020_OPO_visible}. Functionally, this means that for a given $W_\text{p}$, there is a maximum idler wavelength (and hence minimum signal wavelength) for which OPO threshold is reached given the pump power available. This is consistent with the experimental observations in Fig.~\ref{Fig3}, where larger $W_\text{p}$ limits the shortest achievable OPO signal wavelength and longest OPO idler wavelength. 

Conventionally, $Q_0$ in microring resonators is only determined by material absorption and sidewall scattering losses. However, the devices studied here — created in a single fabrication run on the same piece (less than $4$~cm$^2$ area) and with identical microring geometries ($RR$, $H$, and $RW$) — exhibit an order-of-magnitude variation in $Q_0$ at the idler band [Fig.~\ref{Fig4}(c)]. This discrepancy cannot be attributed to fabrication nonuniformity, as the process-induced variations in material and scattering losses are negligible within $4$~cm$^2$ area. The sole distinguishing parameter is the pump/signal waveguide width $W_\text{p}$, which is intentionally varied below the cutoff width for telecom wavelengths. Thus, contrary to initial expectations, where such sub-cutoff $W_\text{p}$ was assumed to negligibly perturb long-wavelength modes (e.g., idler wavelengths $\lambda_i$ $\approx$~$1600$~nm), the observed $Q_0$ degradation implies an unaccounted $\kappa_\text{p}$ arising from the cutoff waveguide, as illustrated in Fig.~\ref{Fig1}(b). 

\begin{figure}[t!]
\centering\includegraphics[width=0.95\linewidth]{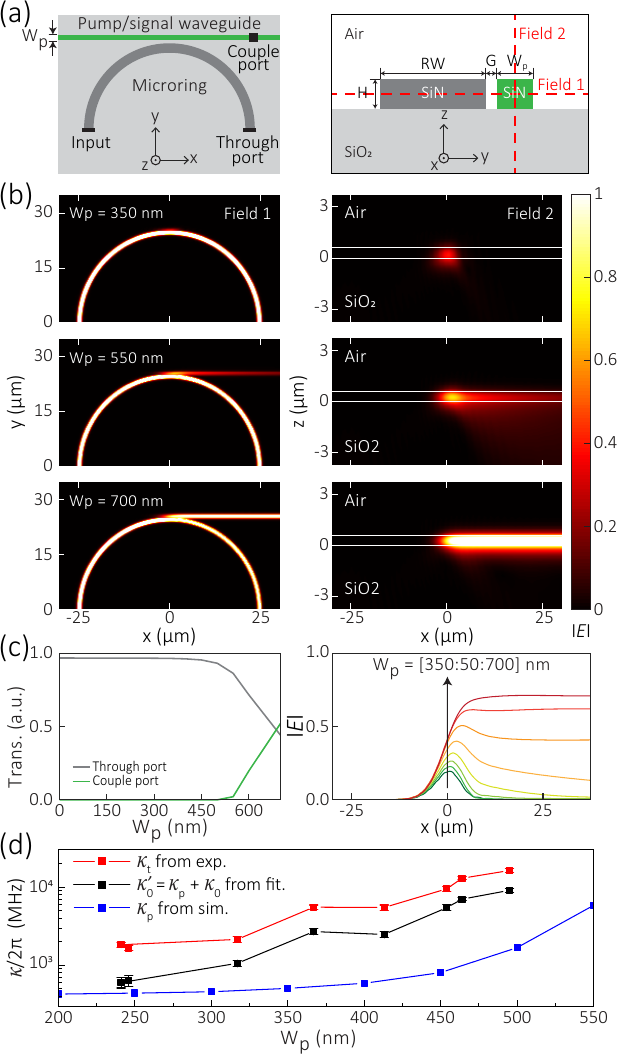}
\caption{\small \textbf{(a)} Schematics of 3D finite-difference time-domain (3D FDTD) simulation configurations from top (left panel) and cross-sectional (right panel) views. \textbf{(b)} Simulated normalized electric field profiles in the microring-pump/signal waveguide coupling region for $W_\text{p}$ = $350$, $550$, and $700$~nm. Left panels show top-view profiles at the plane marked by the red dashed line for field 1, and right panels show cross-sectional profiles along the pump/signal waveguide midplane for field 2, as defined in the right panel of \textbf{(a)}. \textbf{(c)} Left panel shows the transmission at the through port and couple port as the function of $W_\text{p}$. Right panel shows the cut-field lines through the pump/signal waveguide showing the electric field decay lengths. \textbf{(d)} Comparison of $\kappa_\text{p}$ extracted from 3D FDTD simulation with the combination of $\kappa_\text{p}$ and $\kappa_0$ extracted from experiments in Fig.~\ref{Fig4}.}
\label{Fig5}
\end{figure}

We further validate our experimental observation of the reduction in $Q_0$ with increasing $W_\text{p}$ through 3D finite-difference time-domain (FDTD) simulations. We analyze $\kappa_\text{p}$ by focusing on the coupling region of the ring and pump/signal waveguide through the simulation configurations illustrated in Fig.~\ref{Fig5}(a). Detailed simulation setup and methodology is provided in the appendix below. Figure~\ref{Fig5}(b) illustrates the 3D FDTD simulation results for the pump/signal waveguide coupling region at $\lambda=1600$~nm for $W_\text{p}$ of $350$~nm, $550$~nm, and $700$~nm. The top- and cross-sectional- views [field 1 and field 2 in Fig.~\ref{Fig5}(a)] of simulation results are displayed in the left and right panels, respectively. For $W_\text{p}$ = $350$~nm [top panel of Fig.~\ref{Fig5}(b)], the waveguide width lies far below the cutoff, resulting in no observable light coupling into the pump/signal waveguide, and the microring remains unaffected, with nearly all power retained in the through port. As $W_\text{p}$ increases to $550$~nm [middle panel of Fig.~\ref{Fig5}(b)], partial coupling occurs despite the width remaining below the cutoff threshold. Here, light coupled into the pump/signal waveguide radiates and leaks into the substrate along the propagation direction. Although such light is unlikely to reach the chip facet (where it could be measured), it nevertheless results in significant $\kappa_\text{p}$. Finally, at $W_\text{p}$ = $700$~nm [bottom panel of Fig.~\ref{Fig5}(b)], which is close to the cutoff width, efficient coupling is achieved with minimal radiative and leakage losses to the substrate, enabling stable light propagation along the waveguide. 

These observations are quantified in Fig.~\ref{Fig5}(c). The left panel shows transmission at the through and coupled ports as a function of $W_\text{p}$, while the right panel illustrates intensity decay along the pump/signal waveguide. The results clearly suggest three regimes for $W_\text{p}$ in the pump/signal waveguide. Firstly, at considerably smaller $W_\text{p}$ than the nominal cutoff value (i.e., $W_\text{p}$ = $350$~nm), the pump/signal waveguide has little impact on telecom propagation within the ring, resulting in approximately zero power being coupled from the microring to it. Second, in the intermediate regime, near but still below the nominal cutoff value ($W_\text{p}$ ranging from 500~nm to 600~nm), the microring couples telecom light to the pump/signal waveguide, with that coupled light strongly radiating into the substrate as it propagates away from the ring. This light coupled into the pump/signal waveguide results in non-zero $\kappa_\text{p}$ for the telecom (idler) mode. Third, at sufficiently large $W_\text{p}$ with respect to the cutoff value ($W_\text{p}$ = $700$~nm), the microring couples to a propagating waveguide mode with negligible radiative and leakage losses, as typical in a standard resonator-waveguide coupler. 

Parasitic loss induced by the pump/signal waveguide thus depends critically on its width relative to the 700 nm cutoff for $\lambda$ =~$1600$~nm. While conventional designs assume $\kappa_\text{p}$ is negligible for sub-cutoff waveguide widths, our results demonstrate that even in an intermediate regime where $W_\text{p}$ is significantly below cutoff e.g., (at $550$ nm) — where the waveguide cannot guide light over long distances — partial coupling from the microring persists, creating a significant $\kappa_\text{p}$. This coupling to a leaky waveguide mode, visible in the radiative and substrate leakage at $W_\text{p}$~=~$550$~nm [middle panel of Fig.~\ref{Fig5}(b)], highlights the key oversight in assuming sub-cutoff widths universally suppress $\kappa_\text{p}$. To eliminate $\kappa_\text{p}$ entirely, $W_\text{p}$ must be well below the intermediate regime (e.g., $350$~nm in this case), as evidenced by the near-zero intensity decay in the top panel of Fig.~\ref{Fig5}(b). Thus, avoiding the intermediate regime is essential for applications targeting the widest OPO $\nu_\text{i}$-$\nu_\text{s}$ span.

Figure~\ref{Fig5}(d) compares the experimentally extracted and simulated $\kappa_\text{p}$ as a function of $W_\text{p}$. The total loss $\kappa_\text{t}$ (red curve) comprises the microring’s intrinsic loss $\kappa_\text{0}$, the parasitic coupling loss $\kappa_\text{p}$ to the pump/signal waveguide, and the coupling loss to the idler waveguide. The black curve represents the effective intrinsic loss $\kappa_\text{0}'$~=~$\kappa_\text{0}$~+~$\kappa_\text{p}$, derived from quality factor fits [consistent with Fig.~\ref{Fig4}(c)]. The blue curve corresponds to $\kappa_\text{p}$ obtained from 3D FDTD simulations. Despite systematic offsets attributed to unmodeled material absorption and scattering losses in the experiment, both datasets exhibit a monotonic increase in $\kappa_\text{p}$ with $W_\text{p}$, validating the intermediate regime’s role in parasitic coupling. The discrepancy in loss onset bewteen experiment and simulation likely arises from non-idealities not captured in simulations. 

\section{Conclusion}
In this work, we identify and quantify a critical parasitic loss channel in microring resonators arising from cutoff waveguides, a phenomenon previously overlooked in wideband nonlinear photonic designs. While conventional approaches assume that sub-cutoff waveguides negligibly perturb resonator modes, our experiments and simulations reveal a significant intermediate regime where coupling persists, albeit to a waveguide mode that is leaky and has a very short propagation length. In this regime, light partially couples from the microring into the nominally cutoff pump/signal waveguide, radiating into the substrate and introducing $\kappa_\text{p}$ that degrades the intrinsic quality factor of idler modes at long wavelengths by up to an order of magnitude. This effect directly impacts nonlinear processes such as OPO, where elevated losses at the idler wavelength can prevent the system from reaching threshold, thereby limiting spectral coverage. We confirm this behavior in experiments that conclusively demonstrate OPO spans that are dependent on the width of this cutoff waveguide, with stimulated four-wave mixing experiments verifying that phase- and frequency-matching are not the limitation on achievable OPO spans.

Our 3D FDTD simulations elucidate the underlying mechanism: even waveguide widths slightly below the formal cutoff support evanescent coupling sufficient to extract light from the microring, yet insufficient to guide light efficiently, still resulting in radiative leakage. Crucially, $\kappa_\text{p}$ are suppressed only when waveguide widths are designed well below the intermediate regime, as evidenced by the near-zero field decay in such configurations. These findings suggest a revised design guideline for wideband microring devices using air-cladding configures, emphasizing that cutoff compliance alone is inadequate for minimizing $\kappa_\text{p}$. Instead, rigorous avoidance of the intermediate regime is essential to ensure robust pump/signal and idler isolation and to enable nonlinear processes requiring high $Q$ across widely separated spectral bands, such as visible-telecom OPO. Moreover, these design principles are equally critical for other wideband nonlinear photonics processes — like Kerr microcomb generation, SHG, THG, StFWM and FWM-BS — where maintaining high Q at key operative wavelengths is vital for optimal performance. This insight advances the engineering of integrated photonic platforms for applications demanding octave-spanning spectral control.

\medskip
\noindent \textbf{Appendix: Simulation Method}
The geometric parameters in the simulation for the microrings are the same as in the previous experiment with $RR = 25$~{\textmu}m, $RW$ =~$850$~nm, and $H$ =~$600$~nm. $W_\text{p}$ is our sweep parameter, ranging from $200$~nm to $700$~nm with a step size of $50$~nm, while the gap is fixed at $150$~nm. Our simulation setup incorporates three ports: an input port (from the microring) as the mode source, a through port with a mode expansion monitor after the coupling region, and a couple port with a mode expansion monitor at the pump/signal waveguide output. Additionally, two field monitors are positioned strategically -- one in the top view plane at half the waveguide thickness (field 1) and the other in the middle along the length of the pump/signal waveguide (field 2), as shown in Fig.~\ref{Fig5}(a). The entire simulation is enclosed within a region bounded by perfectly matched layers, utilizing a Cartesian mesh grid of $20$~nm. We launch a fundamental $\text{TE}_{0}$ mode at the input port, which propagates in the ring to the through port, with a portion of the power being coupled to the pump/signal waveguide.

To evaluate $\kappa_\text{p}$ in simulations, we utilize the mode expansion monitors and extract the scattering matrices data containing the transmission coefficients. This data is then used to compute the normalized transmission modal powers at both the through and couple ports, as shown in Fig.~\ref{Fig5}(a). However, recording all the power coupled to the pump/signal waveguide at the location of the couple port is challenging due to power radiation and leakage to the substrate caused by the cutoff pump/signal waveguide mode regime. Instead, we calculated the power coupled to the pump/signal waveguide by subtracting the transmitted through power from the input power. Notably, at $W_{p} = 0$ (in the absence of the pump/signal waveguide), we observed that the microring resonator itself introduces additional bending and scattering losses. Thus, to ensure the reliability of our results, we conducted thorough convergence testing. Specifically, regarding the Cartesian mesh grid, we observed little impact on the results (i.e., the transmission spectrum does not uniformly vary with the increase in the mesh grid size). We attribute the resonator’s intrinsic loss to bending and scattering losses from the microring resonator itself, along with a small contribution of loss/uncertainty due to the Cartesian mesh grid. 
Finally, we numerically determined $\kappa_\text{p}$ by subtracting the transmitted through power and the resonator's intrinsic loss from the input power.

\medskip
\noindent \textbf{Data Availability}
The data that supports the plots within this paper and other findings of this study are available from the corresponding author upon reasonable request.

\medskip
\noindent \textbf{Acknowledgements} This work is supported by the DARPA LUMOS and NIST-on-a-chip programs. The authors acknowledge Eric Stanton and Ashutosh Rao for helpful discussions. 

\medskip
\noindent \textbf{Author Contributions} X.L., Y.S., and J.S. carried out the design and simulation of the OPO devices. X.L. and Y.S. carried out the fabrication. Y.S. and J.S. carried out the measurements. D.P. carried out the simulation of parasitic coupling loss. All authors participated in analysis and discussion of results. Y.S. and K.S. wrote the manuscript with the help from others and K.S. supervised the project.


\medskip
\noindent \textbf{Competing Financial Interests} The authors declare no competing financial interests.


\bibliography{idlerQ_CLEO}

\end{document}